\documentclass[prl,aps,showpacs,twocolumn]{revtex4}

\def\be{\begin{equation}}
\def\ee{\end{equation}}

\usepackage{graphicx}
\usepackage{bm}

\begin{document}

\title{
Fermi liquid behavior of the normal phase of a strongly interacting gas of cold atoms}
\author{S. Nascimb\`ene$^1$\footnote{To whom correspondance should be addressed. Present email and address: sylvain.nascimbene@physik.uni-muenchen.de,  Ludwig-Maximilians Universit\"at, Schellingstr. 4, 80799 M\"unchen, Germany}, N. Navon$^1$, S. Pilati$^2$, F. Chevy$^1$,
 S. Giorgini$^3$, A. Georges$^{4,5}$, C. Salomon$^1$}

\affiliation{$^1$ Laboratoire Kastler Brossel, CNRS, UPMC, \'Ecole
Normale Sup\'erieure, 24 rue Lhomond, 75231 Paris, France\\$^2$
Theoretische Physik, ETH Zurich, CH-8093 Zurich, Switzerland
\\$^3$ Dipartimento di Fisica, Universit\`a di Trento and INO-CNR
BEC Center, I-38050 Povo, Trento, Italy\\$^4$ Centre de Physique
Th\'eorique, CNRS, Ecole Polytechnique, route de Saclay, 91128
Palaiseau Cedex, France\\$^5$ Coll\`ege de France, 11 place Marcellin Berthelot, 75005 Paris, France}

\date{\today}

\begin{abstract} We measure the magnetic
susceptibility of a Fermi gas with tunable interactions in the low
temperature limit and compare it to quantum Monte Carlo
calculations. Experiment and theory are in excellent agreement and
fully compatible with the Landau theory of Fermi liquids. We show
that these measurements shed new light on the nature of the
excitations of the normal phase of a strongly interacting Fermi
gas.
\end{abstract}

\pacs{03.75.Ss; 05.30.Fk; 32.80.Pj; 34.50.-s}

\maketitle

In 1956 Landau developed an elegant description of
interacting Fermi systems at low temperature relying on the
existence of long-lived quasiparticles. 
While this Fermi liquid theory (FLT) describes well Helium 3 and
many solid-state materials above the superfluid temperature, there
exist notable exceptions such as underdoped cuprates
\cite{lee2006doping}, where despite tremendous theoretical and
experimental efforts, the nature of the normal phase is not yet
understood. Similarly to high critical temperature
superconductors, the  properties of the normal phase of strongly
correlated atomic fermionic gases and the nature of its
excitations are still debated. This issue was addressed recently
for spin-balanced gases above the superfluid transition, through
the measurement of equations of state
\cite{stewart2006potential,luo2007measurement,horikoshi2010measurement,nascimbene2009eos},
the study of the single-particle excitation spectrum
\cite{gaebler2010observation,perali2010pseudogap} or of spin
fluctuations \cite{Sanner2010}. On the one hand, recent
photoemission spectroscopy experiments near the critical
temperature were interpreted using a pseudogap model
\cite{perali2010pseudogap}. On the other hand, measurement of the
temperature dependence of the specific heat displayed a linear
behavior compatible with Fermi liquid's prediction
\cite{nascimbene2009eos}. All these experimental probes give
access to the properties of the normal phase of the unpolarized
normal phase above the critical temperature $T_c$. This limitation
can be overcome by stabilizing the normal state at $T < T_c$ by
imposing a spin population imbalance in the trapped gas
\cite{partridge2006pap,zwierlein2006fermionic,nascimbene2009pol}
and extrapolating its properties to zero imbalance. Previous works
focused on the highly-polarized limit where minority atoms behave
as impurities: $n_2\ll n_1$, where $n_i$ is the density for
species $i$
\cite{chevy2006upa,lobo2006nsp,combescot2007nsh,shin2008des,prokof'ev08fpb,pilati2008psi,schirotzek2009ofp,nascimbene2009eos,navon2010eos,mora2010normal}.
Here, we interpret the spin imbalance as the application of an
effective magnetic field to the unpolarized normal gas at very low
temperature and using a combination of Monte Carlo simulations and
experimental results, we extract from the equation of state the
magnetic spin response of the normal phase in the limit  $T \ll
T_c$. We show that our results are compatible with a Fermi Liquid
description of the normal phase, and we extract the Fermi liquid
parameters in the universal unitary limit where scattering length
is infinite. The relationship between these parameters and the
properties of low lying excitations of the system allow us to
quantitatively interpret spectroscopic data from
\cite{gaebler2010observation,perali2010pseudogap}.


The polarization dependence of the energy $E$ of the system
directly reflects the presence of spin-singlet dimers in the
sample. Indeed, the presence of a gap in the spin excitation
spectrum implies a linear dependence of the energy $E$ with polarization $p=(N_1-N_2)/(N_1+N_2)$
at low temperature, and hence a zero spin susceptibility. We have performed quantum Monte Carlo simulations of the partially
polarized Fermi gas at $T=0$ in the BEC-BCS crossover. We make use
of the fixed-node diffusion Monte Carlo method that was employed
in earlier studies of polarized Fermi
gases~\cite{lobo2006nsp,pilati2008psi}. The state of
the system is forced to be in the normal phase by imposing the
nodal surface of a many-body wave function incompatible with
off-diagonal long-range order.  A simple way to implement this requirement is by
choosing the trial function of the Jastrow-Slater form
\begin{equation} \psi_T({\bf
R})=\prod_{i,i^\prime}f(r_{ii^\prime}) D(N_1) D(N_2) \;,
\label{psiT} \end{equation} where ${\bf R}=({\bf r}_1,..., {\bf
r}_N)$ is the spatial configuration vector of the $N$ particles
and $D$ denotes the Slater determinant of plane waves in a cubic
box of size $L$ with periodic boundary conditions. The positive Jastrow
correlation term $f(r)$ is determined as described in
Ref.~\cite{lobo2006nsp}: at short distances it corresponds to the
lowest-energy solution of the two-body problem, while it satisfies
the boundary condition on its derivative $f^\prime(r=L/2)=0$.

\begin{figure}[t!]
\includegraphics[width=0.8\linewidth]{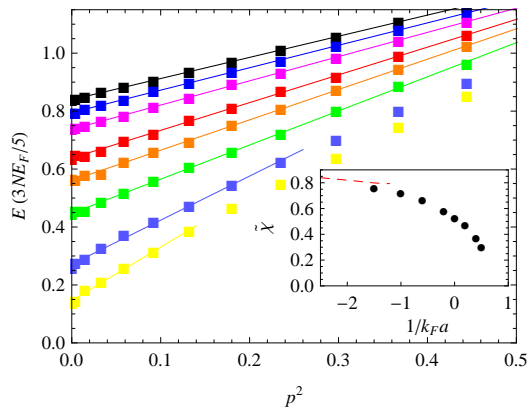}
\vspace{-0.25cm} \caption{Canonical equation of state of a
two-component Fermi gas calculated using quantum Monte Carlo
simulation, for $1/k_Fa=-1.5,-1,-0.6,-0.2,0,0.2,0.4,0.5$ (from top
to bottom). The solid lines are fits of the low-polarization data
with equation (\ref{P2}). Inset: Extracted values of the
susceptibility $\widetilde{\chi}$ as a function of $1/k_Fa$. The
dashed red line is the result of a perturbation expansion valid up
to order $(k_Fa)^2$. \label{Fig_theory}} \end{figure}

 The
results for the canonical equation of state $E(N_1,N_2)$ are
shown in Fig.\ref{Fig_theory}. They are well fitted by the energy
functional \begin{equation} \label{P2}
E(p)=\frac{3}{5}NE_F\left(\xi_N+\frac{5}{9}\widetilde{\chi}^{-1}p^2+\ldots\right),
\end{equation} holding for a spin polarizable system at low temperature, where both $\xi_N$ and the dimensionless spin susceptibility
$\widetilde{\chi}$ (in units of the susceptibility of an ideal
Fermi gas $3n/2E_F$) depend on $1/k_Fa$,  where $k_F=(3\pi^2n)^{2/3}$. The Monte Carlo method
indicates the absence of spin gap, and thus of preformed molecules
in the normal phase for $1/k_Fa\lesssim0.5$. Note that the
extracted values of $\widetilde{\chi}$ reported in the inset of
Fig.\ref{Fig_theory} show a rapid drop  for positive values of $a$
when entering the BEC side of the Feshbach resonance. A likely
explanation is the binding of fermions into spin-singlet pairs for
some positive value of the interaction strength $1/k_Fa$. Monte Carlo calculations for values of $1/k_Fa\ge 0.7$ show that
$E(p)$ is indeed linear rather than quadratic in $p$, indicating the
emergence of a gap. However, pairing fluctuations play a major
role for such values of the coupling and the nodal surface of the
Jastrow-Slater state (see Auxiliary Materials) is no longer sufficient to
enforce the normal phase. This behavior is
reminiscent of  the pairing transition investigated in the framework of BCS theory \cite{combescot2005}, as well as in the normal
phase of the attractive Hubbard model, extrapolated to a
temperature range below the superfluid transition
\cite{capone2002first,toschi2005pairing}, while in our work the
extrapolation is made towards a small spin imbalance.

We now compare these simulations with the grand-canonical equation
of state (EoS) of a homogeneous system obtained experimentally in
ref. \cite{nascimbene2009eos,navon2010eos}.   We prepare a deeply
degenerate mixture of the two lowest internal states of $^6$Li,
held in a cylindrically symmetric hybrid optical/magnetic trap, of
radial (axial) frequency $\omega_r$ ($\omega_z$, respectively).
The bias magnetic field $B_0$ is chosen between 822~G and 981~G,
allowing to tune the strength of interactions $-1<1/k_Fa<0.2$. The final atom
number is $2$ to $10\times10^4$ atoms per spin state, and the gas
temperature is smaller than $0.06\,T_F$, as measured from the fully-polarized
wings of a trapped gas \cite{shin2008pd}.
From dimensional analysis, the EoS of a spin-imbalanced Fermi gas
can be written as \[
P(\mu_1,\mu_2,a)=P_0\left(\mu\right)h\left(\delta=\frac{\hbar}{\sqrt{2m\mu}a},b=\frac{\mu_1-\mu_2}{\mu_1+\mu_2}\right),
\] where $\mu=(\mu_1+\mu_2)/2$ is the mean chemical potential and
$P_0(\mu)$ is the pressure of a non-interacting unpolarized Fermi
gas. $\delta$ is a grand-canonical analog of the interaction
parameter $1/k_Fa$, and $b$ is a dimensionless number proportional
to the `spin-polarizing field' $\mu_1-\mu_2$.


\begin{figure}[t!]
\includegraphics[width=1.05\linewidth]{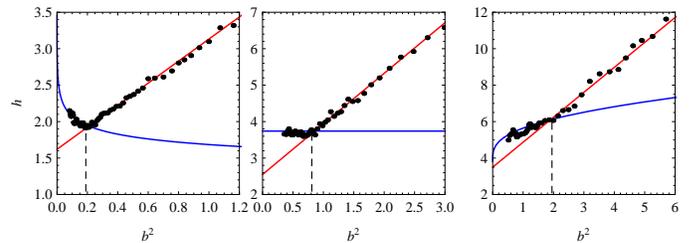}
\vspace{-0.6cm} \caption{Thermodynamic function $h(b)$
measured at different magnetic fields $B_0 = 871, 834, 822$ G. The blue lines correspond to the superfluid
equation of state $h_S(\delta)$ measured in \cite{navon2010eos}.
The red line is a linear fit of the data in the normal phase, $b>
b_c.$ The dashed line indicates the superfluid/normal phase
transition ($b= b_c$). \label{Fig_EoS}}
\end{figure}

\begin{figure}[t!]
\includegraphics[width=0.75\linewidth]{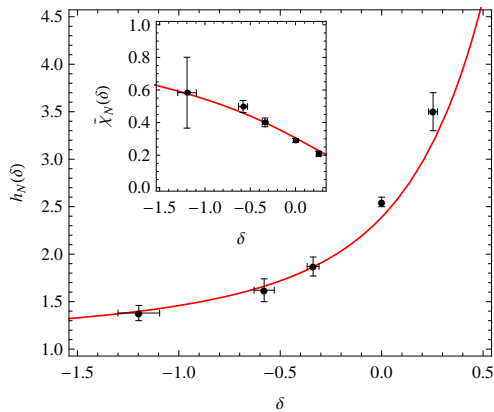}
\vspace{-0.4cm} \caption{Fermi liquid equation of state
extrapolated to a spin-symmetric configuration $h_N(\delta)$. The
black dots are the experimental data, and the red line is
calculated from the Monte Carlo data.  Inset: Grand-canonical
susceptibility $\widetilde{\chi}^{\mathrm{GC}}(\delta)$ of a Fermi gas in
the BEC-BCS crossover. 
\label{Fig_hN}}
\end{figure}

 At all  values of the scattering length addressed in
this work, the equation of state
exhibits a clear discontinuity of its derivative at the critical
field $b_c(\delta)$ (See Fig. \ref{Fig_EoS}), indicating a
first-order phase transition from a superfluid state for $b<b_c$
to a normal state for $b>b_c$ where $h$ is linear in $b^2$.
\cite{zwierlein2006fermionic,navon2010eos}. The equation of state
of the superfluid phase has been discussed in a previous work
\cite{navon2010eos} and we focus here  on the properties of the
normal phase. We write:
\begin{equation}\label{expansion}
h(\delta,b)=h_N(\delta)\left(1+\frac{15}{8}\widetilde{\chi}^{\mathrm{GC}}(\delta)b^2+O(b^4)\right).
\end{equation} $h_N(\delta)$ is the grand-canonical equation of
state in the normal state, extrapolated to a spin-symmetric
configuration. $\widetilde{\chi}^{\mathrm{GC}}(\delta)$ is a grand-canonical
 magnetic susceptibility. For an ideal two-component
Fermi gas, the functions $h_N$ and $\widetilde{\chi}^{GC}$ are equal to 1.
Fitting our data in the normal phase with (\ref{expansion}), we
obtain the parameters $h_N(\delta)$ and
$\widetilde{\chi}^{\mathrm{GC}}(\delta)$ in the BEC-BCS crossover shown in
Fig.\ref{Fig_hN} where we compare their values to the predictions
of the Monte Carlo simulations. To this end, we fit the dependence  with $1/k_Fa$ of the parameters
$\xi_N$ and $\widetilde{\chi}$ determined by Monte Carlo
simulations, and perform a Legendre transform to obtain the
grand-canonical EoS $h_N(\delta)$ of the normal phase and magnetic
susceptibility $\widetilde{\chi}^{\mathrm{GC}}(\delta)$ measured
experimentally. In the investigated parameter range, the agreement
between theory and experiment is excellent.
We also remark that our value for the susceptibility of the normal phase at unitarity is
about twice larger than the value measured in \cite{Sanner2010} on a gas with a 35$\%$ condensate fraction,
confirming a significant suppression of the spin susceptibility in the superfluid phase .


Our findings demonstrate that for $1/k_F a\lesssim 0.5$, the spin
excitations of the system are not gapped in the normal phase which
therefore does not support ``true" molecules. However, a certain
class of theories predicts  a reminiscence of this gap in the form
of a dip in the density of states over a range  $\Delta^*$ around
the Fermi level \cite{chen2005}.  $\Delta^*$ is often called the pseudogap, in
relationship to some features of high-critical temperature
superconductors. These theories predict a departure of $E(p)$ from
its quadratic behavior when the Fermi levels of the two spin
species reach the edges of the dip, $\mu_2-\mu_1\simeq \Delta^*$.
(see Auxiliary Materials). The absence of such an anomaly in Fig.
\ref{Fig_theory},\ref{Fig_EoS}, and in the whole range $-1<1/k_Fa<0.5$ thus suggests that the dip is either extremely  narrow
or very broad: the density of state remains flat over the range of
polarizations and interaction strength studied in our work. For
instance, at unitarity this range covers $0<b^2<3$. If a sizeable
dip existed, then its width cannot be smaller than $\simeq
(\mu_1+\mu_2) \sqrt 3\,\simeq 1.4E_{\rm F}$ where we have used the
unitary equation of state, $\mu=0.41\,E_{\rm F}$
\cite{navon2010eos}. Such a large pseudogap is not compatible with
the photoemission data of \cite{perali2010pseudogap}(See below).
Furthermore, we would expect on physical grounds that $\Delta^*$
becomes smaller on the BCS side of the resonance. This is observed
neither in the experimental data of Fig.\ref{Fig_EoS} nor in the
Quantum Monte Carlo results of Fig.\ref{Fig_theory}.

On the contrary, Landau's  theory of Fermi liquids is fully
compatible with our observations. This theory assumes the
existence of long-lived fermionic excitations above the Fermi
surface. Combining the measurement of the low-temperature
compressibility $\kappa$ and specific heat $C_v$ of
\cite{nascimbene2009eos} with the  data presented here, we can
fully characterize the parameters of the theory at the unitary
limit.  From the magnetic response of the $T=0$ gas, we obtain
here its magnetic susceptibility and another determination of
$\kappa$. The two determinations of $\kappa$ coincide within
5$\%$, showing that the two approaches indeed probe the same Fermi
liquid. From this set of thermodynamic quantities we  derive,
according to Landau's Fermi liquid theory, a complete
characterization of the low-lying excitations of the unitary gas:
besides their effective mass $m^*=1.13\,m$ and Landau parameters
$F_0^s=-0.42$, $F_1^s=0.39$ found in \cite{nascimbene2009eos}, we
recover here $F_0^s=-0.40$ and obtain the new parameter
$F_0^a=m^*/m\widetilde{\chi}(0)^{-1}-1=1.1(1)$. Note that
$F_0^a>0$ corresponds to magnetic correlations which do favor the
singlet configuration.

 \begin{figure*}[ht!]
\includegraphics[width=\linewidth]{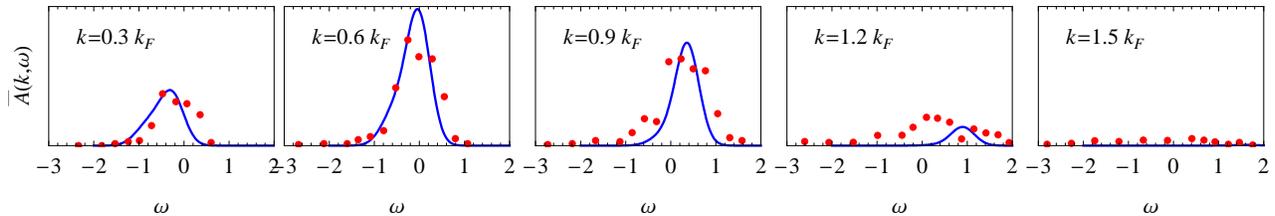}
\vspace{-0.95cm} \caption{Energy distribution data
$\overline{A}(k,\omega)$  from \cite{perali2010pseudogap} (red
dots), compared with the prediction of Fermi liquid theory (blue
lines), for $k/k_F=0.3,0.6,0.9,1.2,1.5$. \label{Fig_spectra}}
\end{figure*}

 We can finally
test FLT on the single-particle photoemission spectrum obtained at
the unitary limit and at the onset of superfluidity from ref.
\cite{perali2010pseudogap}. The experimental signal
$\overline{A}(k,\omega)$ is directly proportional to the spectral
function $A(k,\omega-\mu)$ averaged over the trap that we estimate
using the following procedure: In the vicinity of the Fermi
surface, the dispersion relation of the Fermi liquid
quasi-particles reads
$\hbar\omega_k=\mu+\hbar^2(k^2-k_F^2)/2m^*$ where $m^*=1.13\;m$. Assuming  long-lived quasiparticles,
we approximate $A(k,\omega)$ by $\delta(\omega-\omega_k)$ and
perform the integration over the trap to obtain
$\overline{A}(k,\omega)$ given by
\cite{perali2010pseudogap}:
\begin{equation}\label{EDC}
\overline{A}(k,\omega)=\frac{48k^2}{\pi^2}\int
\mathrm{d}^3r\,\frac{A(k,\omega-\mu(\mathbf{r})/\hbar)}{1+\exp\frac{\hbar\omega-\mu(\mathbf{r})}{k_BT}},
\end{equation} where $\mu(\mathbf{r})$ is the local chemical
potential at position $\mathbf{r}$. In order to calculate the
integrated spectral function $\overline{A}(k,\omega)$ of a Fermi
liquid, we replace the spectral function by $\delta(\omega-\omega_k)$, and perform the integral in (\ref{EDC}). $k_F(r)$ is calculated from the equation
of state of the unitary gas determined in
\cite{nascimbene2009eos}.
The temperature is chosen at the onset of superfluidity
$k_BT/\mu^0=0.32$ \cite{nascimbene2009eos,burovski2006critical}. In order to make a direct comparison with the
experimental data, we finally convolve our result with the
experimental resolution in $\omega$ \cite{perali2010pseudogap}, equal to $0.25E_F/\hbar$ and results for various values of $k$ are shown in
Fig. \ref{Fig_spectra}.

With no free parameter in the theory, FLT well reproduces the
experimental spectra for $\overline{A}(k,\omega)$ in the region
$k<k_F$, with an excellent agreement in the region $0.3\, k_F\le
k\le k_F$ close to the most probable Fermi level in the trap
($\simeq 0.7k_F$) where FLT is expected to be more accurate.
Interestingly, we observe that the width of the peak at $k/k_F=
0.6$ is well reproduced by our model meaning that the broadening
of the line is not limited by the lifetime of
the quasiparticles, but rather by trap inhomogeneity and measurement resolution. Significant deviations between experiment and
FLT appear for
$k>1.1k_F$, far from the most probable Fermi wave-vector. However in this region the energy spectrum signal is very
broad and weak, corresponding to an incoherent background in the
spectral function. Our Fermi liquid description thus accounts
for the coherent part of the excitation spectrum from
\cite{perali2010pseudogap}.

In conclusion we have shown that the magnetic and thermal
responses of the unitary Fermi gas support a description of the
normal phase in terms of Fermi liquid theory despite the fact that
this system exhibits a high critical temperature for
superfluidity. This behaviour is in contrast with underdoped
cuprate high $T_c$ materials displaying anomalous magnetic
susceptibility or pseudogap physics in the normal phase. Recent
quantum oscillation experiments on cuprates in high magnetic
fields, aiming at studying the incipient normal state (somewhat
analogously to the present work) do suggest long-lived
quasiparticles \cite{Taillefer2009}.
 The drop of the susceptibility on the
BEC side of the resonance for $1/k_Fa\gtrsim0.5$ indicates the
appearance of a spin gap in this regime that deserves further
investigations. Finally, the magnetic susceptibility could be a
key observable for characterizing the onset of itinerant
ferromagnetism in a repulsive Fermi gas
\cite{jo2009itinerant,pilati2010itinerant}.

We are grateful to G. Bruun, C. Lobo, P. Massignan, S. Stringari,
T. Giamarchi and R. Combescot for insightful comments. We thank D.
Jin, T. Drake and J. Gaebler for providing us experimental data on
radio-frequency spectroscopy. We acknowledge support from ESF
(FerMix), ANR FABIOLA, R\'egion Ile de France (IFRAF), ERC
Ferlodim, and IUF. S.P. acknowledges support from the Swiss
National Science Foundation, S.P. and A.G. from the Army Research
Office with funding from the DARPA OLE program.

\clearpage
\bigskip
\noindent\textbf{AUXILIARY MATERIAL}\\

 \setcounter{figure}{0} \setcounter{table}{0}
 \makeatletter \renewcommand{\thefigure}{A\@arabic\c@figure} \renewcommand{\thetable}{A\@arabic\c@table} \makeatother

\renewcommand{\bibnumfmt}[1]{[A#1]}

\noindent\textbf{Measurement of the equation of state}
We recall here the procedure
used to measure the equation of state $P(\mu_1,\mu_2,a)$, that was
already employed in \cite{navon2010eosA} after the method proposed by
\cite{ho2009opdtq}.  We prepare a deeply
degenerate mixture of the two lowest internal states of $^6$Li,
held in a cylindrically symmetric hybrid optical/magnetic trap, of
radial (axial) frequency $\omega_r$ ($\omega_z$, respectively).
The bias magnetic field $B_0$ is chosen between 822~G and 981~G,
allowing to tune the strength of interactions. The final atom
number is $2$ to $10\times10^4$ atoms per spin state, and the gas
temperature is $0.03(3)T_F$, as measured from the fully-polarized
wings of a trapped gas \cite{shin2008pd}. As shown in
\cite{ho2009opdtq,nascimbene2009eosA,navon2010eosA}, the local gas
pressure along the $z$ axis can directly be obtained from its
\emph{in situ} image. In the framework of local density
approximation, this provides the grand-canonical equation of state
$P(\mu_1,\mu_2,a)$ at the local chemical potentials
$\mu_{iz}=\mu_i^0-\frac{1}{2}m\omega_z^2z^2$, where $\mu_i^0$ is
the global chemical potential for species $i$. The global chemical
potential $\mu_1^0$ for the majority species is directly obtained
from the Thomas-Fermi radius $R_1$ of the fully polarized phase,
according to $\mu_1^0=\frac{1}{2}m\omega_z^2R_1^2$. Similarly to
\cite{navon2010eosA}, we obtain the global chemical potential
$\mu_2^0$ by imposing that, at the outer radius $R_2$ of the
minority species, the chemical potential ratio $\mu_2/\mu_1$ is
given by the resolution of the impurity problem
\cite{chevy2006upaA,lobo2006nspA,combescot2007nshA,prokof'ev08fpbA,pilati2008psiA,schirotzek2009ofpA}.

\bigskip
\noindent\textbf{Fixed-Node
Monte Carlo simulation}
The Hamiltonian of the
$N=N_1+N_2$ atoms of the two species is given by \begin{equation}
H=-\frac{\hbar^2}{2m}\left( \sum_{i=1}^{N_1}\nabla^2_i +
\sum_{i^\prime=1}^{N_2}\nabla^2_{i^\prime}\right)
+\sum_{i,i^\prime}V(r_{ii^\prime}) \;, \label{hamiltonian}
\end{equation} where $i,j,...$ and $i^\prime,j^\prime,...$ label,
respectively, majority and minority fermions. We model
interspecies interatomic interactions using an attractive square
well potential: $V(r)=-V_0$ if $r<R_0$ and zero otherwise
($V_0>0$). The short range $R_0$ is fixed by the condition
$nR_0^3=10^{-6}$, where $n=n_1+n_2$ is the total atom density. The
depth $V_0$ is instead chosen as to give the proper value of the
scattering length $a$ along the BEC-BCS crossover. We consider a
system with fixed total number of particles ($N=66$) in a fixed
volume $V=L^3$.

Finite-size effects have been reduced using the technique described in \cite{pilati2010itinerantA}.

\noindent\textbf{Gap and spin susceptibility}
Let us consider a system containing $N_1$ spin up and $N_2$  spin down particles.
 We define $M=N_1-N_2$ and $N=N_1+N_2$ the polarization and the total
  atom number and we note $E(N,M)$ the energy of the system. If one assumes that the energy can be expanded
  in $M$ then by symmetry the linear term vanishes and one gets $E(N,M)=E(N,0)+M^2/2\chi+...$. With this definition,
  $\chi$ is then the spin susceptibility of the system. Indeed, adding a magnetic field $h$ contributes to a $-hM$
  term to the energy and we immediately see that the energy minimum is shifted from $M=0$ to $M=\chi h$.

This argument is no longer true in the case of a gapped system.
Indeed, polarizing a spin balanced system costs the binding energy
of the broken pairs. This definition applies to any system
composed of spin-singlet dimers, from a fermionic superfluid
composed of Cooper pairs, or a pure gas of uncondensed molecules,
and leads to the following leading order expansion
$$
E(N,M)=E(N,0)+|M|\Delta+...
$$
To evaluate the spin susceptibility, we add as above a magnetic field $h$ changing the energy into $E-h M$.
We see that for $h\not =0$, the potential is tilted but the energy minimum stays located at $M=0$
(as long as $|h|<\Delta$ corresponding to the Pauli limit pointed out by Clogston and Chandrasekhar in the case of superconductors \cite{Clogston1962A,Chandrasekhar1962A}).

\begin{figure}[h]
\includegraphics[width=0.8\columnwidth]{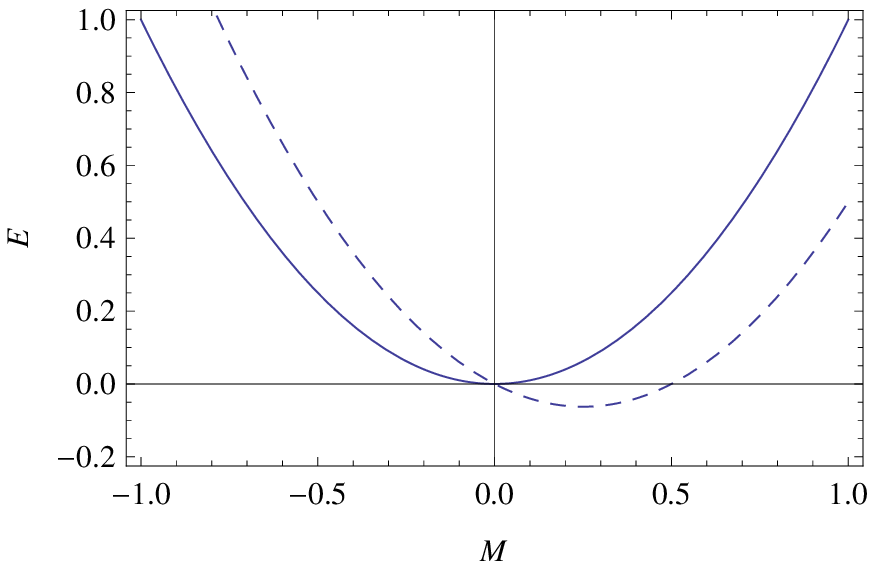}
\includegraphics[width=0.8\columnwidth]{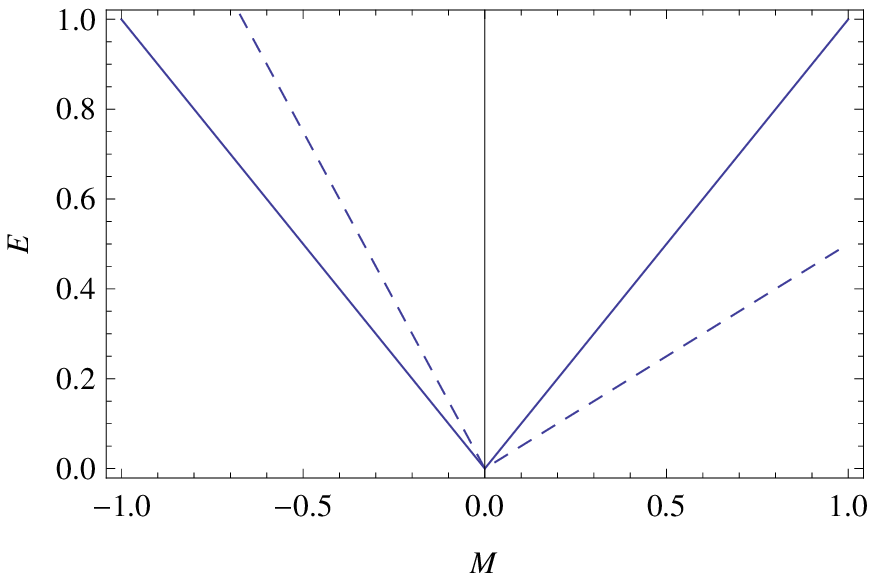}
\caption{Dependence of energy $E$ with spin population imbalance
$M$ for gapless (top) and gapped (bottom) systems. Top. Full line:
The dependence with spin imbalance is quadratic and the curvature
is equal to the inverse of the spin susceptibility $\chi$.
Dashed line: In the presence of a spin polarizing field $h$, the
energy minimum is shifted to $M=\chi h$. Bottom: gapped system.
Full line: Energy in the absence of external spin polarizing
field. The slope is equal to the gap $\Delta$. Dashed line: in the
presence of a spin polarizing field, the energy profile is tilted
but the minimum remains located at $M=0$.} \label{Fig_Energy}
\end{figure}

\begin{figure}[h!t!]
\includegraphics[width=\columnwidth]{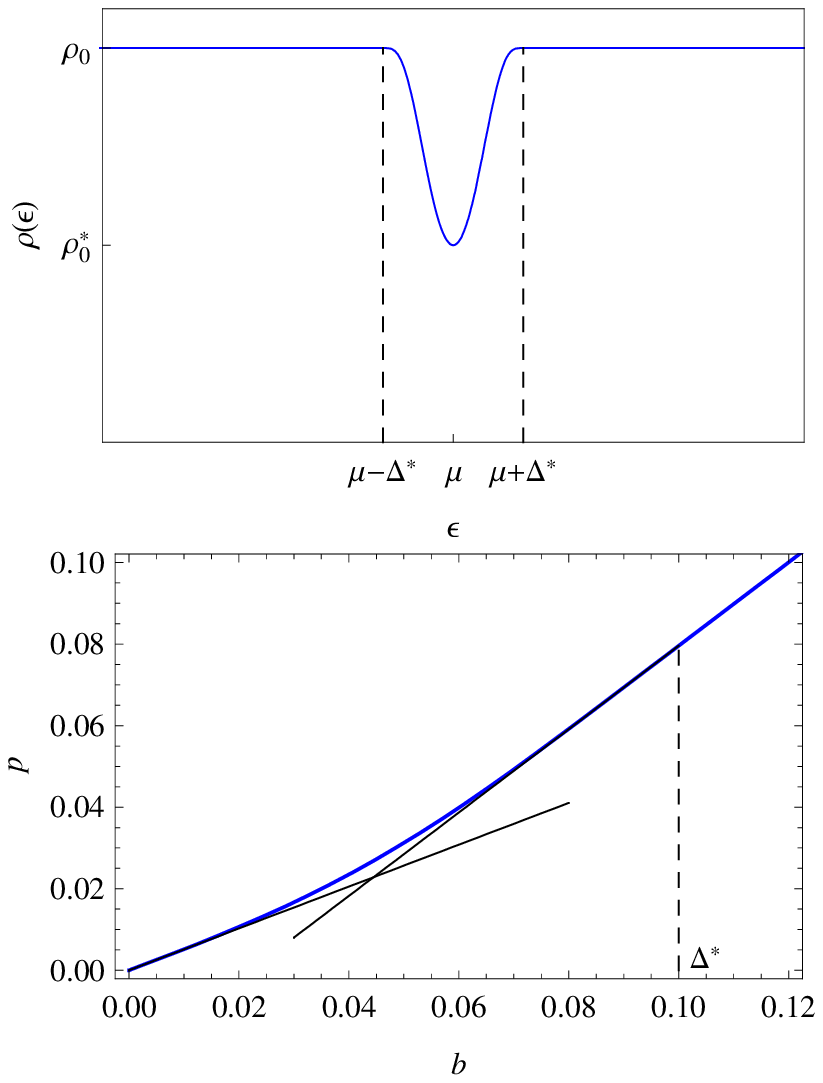}
\caption{Top: sketch of the density of state in a pseudogap model. A footprint of the molecular state appears as a dip of width $\Delta^*$ and depth $\rho_0-\rho_0^*$ in the density of state. Bottom: Polarization $p$ as a function of the magnetic field $b$, with $\rho_0^*=0.5\rho_0$ and $\Delta^*=0.1\mu$. At low imbalance, the Fermi levels of the two spin species lie inside the dip. This results in a depletion of spin excitations and a reduction of the spin susceptibility.}
\label{Fig:Pseudo}
\end{figure}

\bigskip \noindent\textbf{Thermodynamic signature of the pseudogap} The pseudogap phenomenon can be defined as a dip in the density
of state $\rho (\varepsilon)$ close to $E_F$ reminiscent of the
true cancellation of $\rho$ inside the superfluid gap. We note
$\Delta^*$ the width of the dip, and for the sake of simplicity we
assume that $\Delta^*\ll k_B T_F$. In a simple model where one
assumes that the excitations of the system are described by the
Fermi-Dirac distribution, one can show that the spin
susceptibility of the system is equal to the density of
state at the Fermi level. For small imbalances, the Fermi levels
of the two spin states lie within the dip. The spin susceptibility
is thus $\chi\simeq\rho_0^*$. When the imbalance is larger, the
two Fermi surfaces are outside the dip
 (when $E_{F1}-E_{F2}\gg \Delta^*$), and the pseudogap excitations do not contribute anymore to the thermodynamics of the system.
 In this case, the spin susceptibility is given by $\chi=\rho_0>\rho_0^*$ (see Fig. \ref{Fig:Pseudo}).

\begin{figure}[h!t!]
\includegraphics[width=\linewidth]{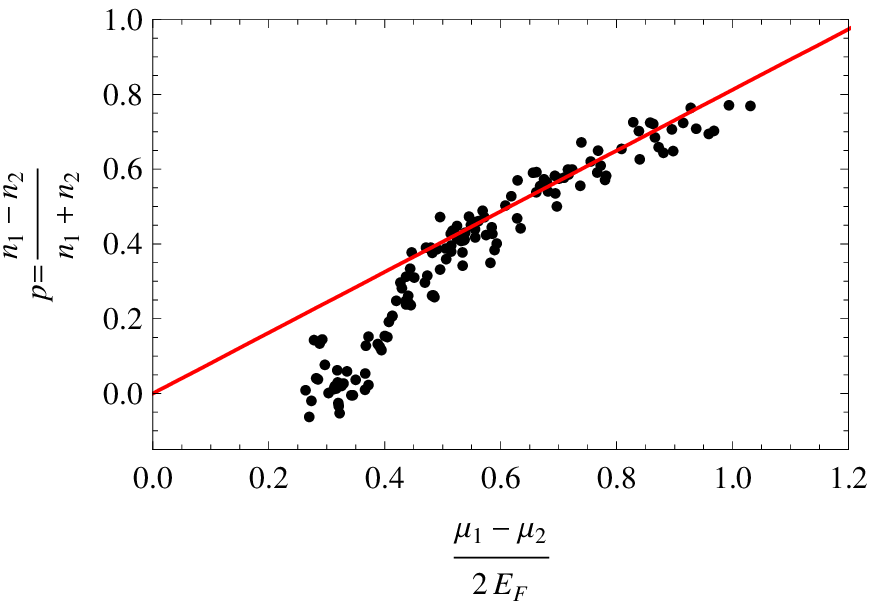}
\vspace{-0.6cm} \caption{Polarization $p$ of the unitary Fermi gas
as a function of the magnetic field $\mu_1-\mu_2$, normalized to
the Fermi energy $E_F=\hbar^2/2m(3\pi^2n)^{2/3}$. The red line
corresponds to the relation $p=\frac{3}{2}\widetilde{\chi}(\mu_1-\mu_2)/2E_F$, with $\chi=0.54$.
\label{Fig_unitary_canonical}}
\end{figure}

\bigskip \noindent\textbf{Polarizability of the
unitary Fermi gas} It is interesting to express our data for the unitary
Fermi gas in the usual variables of condensed matter physics,
namely the polarization $p=(n_1-n_2)/(n_1+n_2)$ as a function of
$(\mu_1-\mu_2)/2E_F$, where $E_F$ is the Fermi energy. These
quantities are calculated from the thermodynamic function at
unitarity $h_0(b)=h(\delta=0,b)$ according to \[
p=\frac{h_0'(b)}{\frac{5}{2}h_0(b)-b\,h_0'(b)},\quad\frac{\mu_1-\mu_2}{2E_F}=\frac{b}{(h_0(b)-\frac{2}{5}b\,h_0'(b))^{2/3}}.
\] This requires to take the derivative of our experimental data,
which decreases the signal-to-noise ratio. We obtain the data
plotted in Fig.\ref{Fig_unitary_canonical}. In the superfluid
phase the polarization remains equal to 0 and jump to $p\simeq0.4$
at the superfluid/normal transition (for
$\mu_1-\mu_2\simeq0.4\cdot2E_F$). The polarization then increases
linearly with the magnetic field over a large polarization range,
according to $p=\frac{3}{2}\widetilde{\chi}(\mu_1-\mu_2)/2E_F$, with
$\widetilde{\chi}=0.54$.

\end{document}